\documentclass[12pt,preprint]{aastex}

\usepackage{lscape}

\shorttitle{SFPR techniques applied to Space VLBI}
\shortauthors{Rioja etal.}

\begin{document}

\title{Exploration of SFPR techniques for astrometry and observations
  of weak sources with high frequency Space VLBI.}

\author{M. Rioja\altaffilmark{1,2}, R. Dodson\altaffilmark{1}, 
J. Malarecki\altaffilmark{1} and Y. Asaki\altaffilmark{3,4}}
\affil{$^1$ ICRAR, UWA, Perth, Australia}
\altaffiltext{2}{On secondment Observatorio Astron\'omico Nacional
  (OAN), Spain.} 
\affil{$^3$ Institute of Space and Astronautical Science,
   3-1-1 Yoshinodai, Chuou, Sagamihara, Kanagawa 252-5210, Japan}
\affil{$^4$    Department of Space and Astronautical Science,
   School of Physical Sciences,
   The Graduate University for Advanced Studies (SOKENDAI),
   3-1-1 Yoshinodai, Chuou, Sagamihara, Kanagawa 252-5210, Japan}

\email{maria.rioja@icrar.org}



\begin{abstract}

Space Very-Long-Baseline-Interferometry (S-VLBI) observations at high frequencies
hold the prospect of achieving the highest angular resolutions and astrometric
accuracies, resulting from the long baselines between ground and satellite telescopes.  
Nevertheless, space-specific issues, such as limited accuracy in the satellite 
orbit reconstruction and constraints on the satellite antenna pointing
operations, limit the application of conventional phase
referencing. 
We investigate the feasibility of an alternative
technique, {\sc source frequency phase referencing} (SFPR), 
to the S-VLBI domain. 
With these investigations we aim to contribute to 
the design of the next-generation of S-VLBI
missions.
We have used both analytical and simulation studies to
characterize the performance of SFPR in S-VLBI observations,
applied to astrometry and increased coherence time, and compared these
to
results obtained using conventional phase referencing.  
The observing configurations use the specifications of the ASTRO-G
mission for their starting point.
Our results show that the SFPR technique enables astrometry at 43 GHz,
using alternating observations with 
22 GHz, regardless of the orbit errors, for most weathers and
under a wide variety of conditions. The same applies to the increased
coherence time for the detection of weak sources.  Our studies show
that the capability to carry out simultaneous dual frequency observations 
enables the application to higher frequencies, and a general improvement 
of the performance in all cases,
hence we recommend its consideration for S-VLBI programs.

\end{abstract}

\section{Introduction}\label{sec:intro}

The pursuit of ever higher angular resolution is the driver
pushing VLBI observations into higher radio frequency domains, and
increasingly larger telescope separations.  Advances on both
fronts have allowed the mapping of
the structure of astronomical objects, such as distant active galactic
nuclei (AGN), at steadily increasing resolution.  A major step
forward is expected from the combination of observations with the
longest baselines and at high frequencies, that is, with the
next-generation of S-VLBI. 
This will target a range of fundamental
physical problems, such as measuring the properties of accretion
disks in super-massive black holes, amongst others
(see http://www.vsop.isas.jaxa.jp/vsop2e/science, as well as
\citet{takahashi_04,takahashi_11}).
Suitable astrometric accuracy and the high sensitivity required to
measure these properties are important tools to provide a path to
those ends.  This paper is concerned with developments of calibration
techniques that enable these outcomes, in the light of current
and future space missions.

Joint observations between ground radio telescopes and the 
Japanese satellite HALCA, launched in 1997 \citep{hirax_00}, demonstrated 
the feasibility of S-VLBI. HALCA operated at 1.6/5 GHz, 
mainly for imaging purposes using self-calibration techniques (e.g. \citet{dod_08}), 
and some limited astrometry \citep{porcas_00,guirado_01}.
Following this success a number of S-VLBI astronomical projects are
currently under development,
such as ASTRO-G \citep{tsuboi_09}, RadioAstron \citep{kardashev_97},
Millimetron \citep{wild_09}. Their mission specifications comprise a
range of orbit apogees from 25,000 to 350,000 km, with most frequencies
in the cm and mm range,
and satellite antenna diameters in the range 9 to 12 meters.
For more detailed information on these missions see the references listed.

The long baselines involved in S-VLBI hold the potential of
achieving the highest spatial resolution at any particular
frequency. Nevertheless, the requirements for the astrometric capability and enhanced
sensitivity that are achieved with conventional phase referencing
using ground arrays are difficult to meet with a
spacecraft,
particularly at the higher frequencies ($> 22$ GHz).  For example, the
astrometric capability strongly depends on the Orbit Determination
Discrepancy at Apogee (ODDA) and requires cm-level orbit
reconstruction, which is extremely difficult to realize using conventional 
range and range-rate satellite tracking techniques \citep{asaki_orbit}.  

Conventional phase referencing analysis \citep{alef_88,beasley_95}
routinely achieves high precision astrometric measurements using 
observations with ground arrays in the range between 2 and 43 GHz.
Its successful implementation is strongly dependent on the existence of
a nearby, compact and strong calibrator 
for alternating observations with the target source, 
and having accurate a-priori models 
for the source and antenna positions, and atmospheric
effects, amongst others. 
The rapid source-switching cycles required to compensate for the
tropospheric fluctuations at high frequencies, the scarcity of
suitable calibrator sources and the constraints on a priori models
pose an insurmountable limitation to the application of this technique
beyond 43 GHz.

In theory, phase referencing
techniques can be applied both to space and
ground baselines.  In practice, the analysis of HALCA data
showed that the calibration of S-VLBI observations
involves additional difficulties arising from the lower correlated
source flux densities at the higher resolution of space baselines,
relatively poor sensitivity achievable with small orbiting antennas,
technical difficulties of rapid pointing changes for 10-m class 
deployable antennas in space and large geometric delay errors 
introduced by uncertainties in the spacecraft orbit preventing 
long integration times, and astrometry \citep{porcas_00}.

Asaki and collaborators (2007) ({\it hereafter} A07) have
carried out a comprehensive study on the feasibility of conventional
phase referencing observations with ASTRO-G, under a range of
different weather conditions, such as source separations and orbit
determination accuracy of the satellite, among other parameters. Their
simulations conclude that astrometrical observations are expected to
achieve a good performance at 8.4 GHz, while at higher frequencies the
best possible weather would be required and the probability of finding
suitable calibrators, particularly at 43 GHz, is greatly reduced.
In all cases, cm-level orbit reconstruction is required and 
additional instruments for precise orbit determination
would be needed on board, such as Global Navigation Satellite System navigation, 
Satellite Laser Ranging, etc. as described in \citet{asaki_orbit} and \citet{wu_gps}.

We propose an alternative phase calibration strategy that widens the
astrometric capability and enhances the sensitivity of S-VLBI, by
addressing the space-specific and high frequency issues mentioned
above.  By doing that, it allows the application to many targets and
frequencies beyond the limits of conventional phase referencing.
The {\sc source frequency phase referencing} (SFPR {\it hereafter})
technique consists of using observations at another (lower) frequency,
plus another source to calibrate the target source observations at a
higher frequency. 
The direct astrometric outcome of SFPR observations is high precision
{\it bona fide} astrometry between frequencies,
of interest in studies where the spatial alignment of emission, continuum 
or spectral line, at multiple frequencies is crucial;
when combined with conventional phase referencing (PR {\it hereafter}) 
at the lower frequency,
this enables relative astrometry with respect to an external reference 
at the higher frequency, for proper motion, parallax, and other such studies.
The basis of the SFPR
strategy is presented in detail in \citet{rioja_11} (RD11 {\it hereafter}), 
and \citet{dodson_m31}, along with an error analysis, and an
empirical demonstration with observations using the VLBA ground array
at 43 and 86 GHz.
This paper is concerned with its application to S-VLBI, and is a 
development of the VLBA Memo by \citet{rioja_m32}.
In Section 2 we focus on the advantages and suitability of the SFPR
calibration method to provide astrometry and increased coherence
for mm-wavelength S-VLBI observations.
In Section 3 we present the results from analytical and simulation studies
which use the specifications of the
ASTRO-G mission as a starting point. Section 4 aims at
a more general discussion of S-VLBI, and provides suggestions for
design improvements.

\section{The SFPR technique and feasibility studies for Space VLBI}
\label{sec:method}

\subsection{The SFPR technique} \label{sec:m_technique}

The two-step SFPR astrometric calibration approach relies on {\it fast
  frequency switching}, or ideally simultaneous dual frequency
observations, combined with {\it slow source switching} observations,
between two frequencies ($\nu^{high}$ and $\nu^{low}$) and two sources
({\it A} and {\it B}), respectively.  The former step alone provides a
method to compensate for the non-dispersive errors in the tropospheric
excess delay model,  
hence effectively increasing the sensitivity of S-VLBI in the high
frequency regime, by increasing the coherence time; the orbit
determination errors being non-dispersive are compensated in this step
as well.
When combined with the latter step it enables astrometric capability
with S-VLBI irrespective of the uncertainties in the orbit
reconstruction, which set an insurmountable limiting factor with
PR techniques at high frequencies.  Also, since the constraints on a
suitable calibrator source are much less severe than in conventional PR, it
enables S-VLBI astrometry of many targets at 43 GHz and higher
frequencies.

A detailed presentation of the basics of the SFPR technique 
with ground arrays can be found in RD11.  
In order to facilitate the reading of this paper we include an extract
of the formulae used in RD11, 
with emphasis on aspects which are specific for S-VLBI.
Following standard nomenclature, the residual phase error
values for observations of the target source ($A$) at the target
frequency, $\phi^{high}_{A}$, are shown as a compound of geometric,
tropospheric, ionospheric, instrumental, structural 
and thermal noise residual terms:

\begin{equation}
\phi^{{\rm high}}_{{\rm A}} = \phi^{{\rm high}}_{{\rm A,geo}} 
 + \phi^{{\rm high}}_{{\rm A,tro}} +\phi^{{\rm high}}_{{\rm A,ion}}
 +\phi^{{\rm high}}_{{\rm A,inst}} 
 +\phi^{{\rm high}}_{{\rm A,str}}
 +\phi^{{\rm high}}_{{\rm
     A,thermal}} + 2\pi n^{{\rm high}}_{{\rm A}}, \hspace*{1cm}
 n^{{\rm high}}_{\rm A} \in {\rm integer},
\end{equation}

\noindent
where $2\pi n^{{\rm high}}_{\rm A}$ stands for the intrinsic phase
ambiguity term. 
$\phi^{{\rm high}}_{{\rm A,str}}$  corresponds to the structure
contribution and can be calculated with respect to a feature in the
maps. By choosing the ``core'' component as the phase center the $\phi^{{\rm
    high}}_{{\rm A,geo}}$ term refers to the position of this
``core''. This is the criterion adopted in this paper.

A similar expression to equation (1) holds for the residual phases
$\phi^{low}_{A}$ from observations at $\nu^{low}$, the reference
frequency. These are analyzed using self-calibration and hybrid
imaging techniques so that $\phi_{\rm str}$ is compensated. The
resultant antenna-based corrections are linearly interpolated to the
times when the $\nu^{high}$ frequency is observed ($\tilde
\phi^{low}_{A,self-cal}$), scaled by the frequency ratio $R$ (with
$R=\frac{\nu^{high}}{\nu^{low}}$), and applied as the calibration for
the observed phases at $\nu^{high}$, the target frequency, in equation
(1).  We name this step as {\sc frequency phase transfer} ({\sc fpt}).
This calibration strategy results in quasi perfect compensation of
non-dispersive residual phase model errors which scale linearly with
frequency, such as the tropospheric and geometric contributions, in
equation (1). Then, the residual tropospheric contribution is given
by:

\begin{displaymath}
\phi_{{\rm A,tro}}^{{\rm high}} - R\,.\,\tilde \phi_{{\rm
    A,tro}}^{{\rm low}} 
= \Delta_{{\rm i,T^{\nu}_{\rm swt}}} ,
\end{displaymath}

\noindent
where $\Delta_{{\rm i,T^\nu_{\rm swt}}}$ stands for the interpolation
errors arising from using a frequency switching cycle $T^\nu_{{\rm
    swt}}$,
which corresponds to the
elapsed time between midpoints of two consecutive scans at the same
frequency. The propagation in the SFPR analysis is addressed in latter sections.
This term can be reduced by selecting a fast frequency switching
cycle, matching the properties of the tropospheric fluctuations.
Frequency switching is easier than source
switching for telescopes in general. The ideal configuration consists of using simultaneous dual
frequency observations, for which neither switching nor interpolation
is required and $\Delta_{{\rm i,T^\nu_{\rm swt}}}$=0.

As for the geometric compensation, it is given by: 

\begin{equation}
\phi^{high}_{A,geo}  - R\,.\,\tilde\phi^{low}_{A,geo}  = 
2\pi \,\vec{D_{\lambda}}\,.\,\vec{\theta_{A}} + O (\vec{\Delta
D_{\lambda}}\,.\,\vec{\theta_A})
\approx 2\pi \,\vec{D_{\lambda}}\,.\,\vec{\theta_{A}}       
\end{equation}

\noindent
where $\vec{D_{\lambda}}$ is the baseline vector in units of
wavelengths (for $\nu^{high}$), and $\vec{\theta}_A$ is the target
source position shift between the two observed frequencies. This we
refer to as ``core shift'' by extension of the core shift phenomena in
AGNs but also applicable to any spatial spectral shift, independent of
its origin.  In general, for sources whose VLBI position is frequency
dependent, the geometric contribution has a 24-hour sinusoidal term
whose amplitude depends on $\vec{\theta}_A$.  For completeness we
include an extra contribution proportional to the scalar product of
the antenna position error (or satellite orbit error) and the ``core
shift'' vectors. The effect of the latter extra term is negligible and
can be completely ignored given the likely orbit errors, or any other
VLBI antenna position errors, and the expected typical values for
``core shifts''.

Then, the resultant tropospheric and geometric error-free 
residuals, so called {\sc FPT}-calibrated target phases,
$\phi_{\rm A}^{{\rm FPT}}$, are:

\begin{center}
\begin{eqnarray}
\phi_{\rm A}^{{\rm FPT}} = \phi_{\rm A}^{{\rm high}} - R\,.\,\tilde
\phi_{{\rm A,self-cal}}^{{\rm low}} = \phi_{{\rm A,str}}^{{\rm high}}
+ 2\pi \,\vec{D_{\lambda}}\,.\,\vec{\theta_{A}}
+ (\phi_{{\rm A,ion}}^{{\rm high}} - R\,.\,\tilde \phi_{{\rm
    A,ion}}^{{\rm low}})  \nonumber \\ 
+ (\phi_{{\rm A,inst}}^{{\rm high}} -
R\,.\,\tilde \phi_{{\rm A,inst}}^{{\rm low}}) 
+ \Delta_{{\rm i,T^\nu_{\rm swt}}}
\end{eqnarray}
\end{center}

For simplicity we have omitted the noise contribution, and the $2\pi$ phase
ambiguity term in Equation 3 which, provided $R$ is an integer value,
just adds an unknown number of whole turns and is irrelevant for
the analysis.  The importance of having an integer ratio between the
frequencies involved in SFPR calibration is discussed in RD11;
nevertheless, successful analysis using non-integer frequency ratios
has been demonstrated \citep{rioja_05,dodson_eavw_11}.
Also, we take $\phi_{{\rm A,str}}^{{\rm low}}=0$, either because 
the structure at $\nu^{low}$ has been imaged and corrected for, 
or it is a compact source.
Note that the compensation of the tropospheric short time scale phase
variations results in longer coherence times at the higher frequencies
$\nu^{high}$, which enables the detection of weaker sources
irrespective of the orbit errors. This is of special interest for
S-VLBI given the limitations on the size of a satellite antenna
and the long baselines. However the remaining dispersive residual
phase contributions prevent astrometry.

Interleaving observations of another source B, following 
the same strategy as for the target source A, offers a way to calibrate
the remaining dispersive residual terms in Equation 3.
Note that since the remaining ionospheric and instrumental terms show
long-scale temporal variations, a slow source switching of several
minutes, along with large angular separations of several degrees is
feasible. Provided that suitable switching times are used during the observations,
the resultant {\sc sfpr}-visibility phases, $\phi^{{\rm
SFPR}}_{{\rm A}}$, are free of the long scale drift terms shown in Equation
3, as: 

\begin{equation}
  \phi^{{\rm SFPR}}_{{\rm A}} = \phi_{{\rm A,str}}^{{\rm high}} + 2
  \pi \vec{D}_{\lambda} \, . \, (\vec{\theta}_{{\rm A}}-
  \vec{\theta}_{{\rm B}}) + \Delta_{{\rm i,T^\nu_{\rm swt}}} +
  \Delta_{{\rm i,T_{\rm swt}}} 
\end{equation}

\noindent
where $\Delta_{{\rm i,T_{\rm swt}}}$ stands for the interpolation
errors arising from using a source switching cycle ${\rm T}_{{\rm
    swt}}$, which corresponds to the
elapsed time between midpoints of two consecutive blocks of scans 
on the same source. The propagation of these interpolation errors in the SFPR analysis 
is addressed in latter sections.
The structure contributions for source B at both frequencies are 
calculated from the corresponding self-calibration maps, as explained above. 
The {\sc sfpr}-calibrated phases are free of
geometric, tropospheric, ionospheric and instrumental corruption while keeping
the chromatic astrometry signature of frequency dependent position.
It is interesting to note that the
``core-shift'' functional dependence in Equation 4 is identical to
that for the pair angular separation in conventional PR, although this
method cannot be applied at high frequencies.  
Finally, a Fourier transformation of the visibilities, without further
phase calibration, results in a {\sc sfpr}-map of the brightness
distribution of the target source ($A$) at $\nu^{high}$ frequency, and
where the offset of the peak with respect to the center of the map is
astrometrically significant: i.e. a measurement of the relative ``core
shift'' between the observed frequencies ($\nu^{high}$ and $\nu^{low}$), for
both sources ({\it A} and {\it B}).

\subsection {S-VLBI satellites: ASTRO-G and RadioAstron} \label{sec:m_astrog}

Table \ref{tab:ag} lists the basic parameters for ASTRO-G and
RadioAstron, the best current examples of S-VLBI missions.
A full description for the ASTRO-G spacecraft can be found in
\citet{tsuboi_09} and for RadioAstron in 
\citet{kardashev_97}.  

ASTRO-G is equipped with three frequency horns, for X, K and Q bands,
each with a slight pointing offset in the optics design.  ASTRO-G therefore
needs to alter the satellite body attitude in order to switch between
observing frequency bands for any particular source.  Rapid attitude
change can be achieved with the powerful attitude control actuators
designed to allow for source switching
over angles of $\sim \,3\,$ degrees with a switching period of 1
minute.  It is expected that for the very small switching angles
required in frequency switching even shorter cycle times could be
achieved.  Integer frequency ratios exist between the X- and Q-bands
and K- and Q-bands of 5 and 2, respectively. For the analytical
studies we have also assumed that SFPR observations between X- and
K-band are possible, even though an integer ratio does not exist.

RadioAstron was launched on June 18th 2011 on a Zenit-3M launcher and
at the time of writing this paper is in checkout phase.
It is equipped with 4 frequency receivers in a concentric arrangement,
which allows for simultaneous observations of the bands. The frequencies are
listed in table \ref{tab:ag} and span 1-meter to 1-cm.  
The maximum antenna slew speed is 3-min/$^\circ$ \citep{rauh}.
Based on these specifications, SFPR observations at C/K-bands, 
with integer frequency ratios of 4 and 5, would be feasible.
This is discussed later in this paper, in the context of
the large orbit and orbit uncertainties of RadioAstron.

\begin{table}
\begin{center}
  \caption{Summary of the specifications of two S-VLBI missions. Left)
    the ASTRO-G satellite and right) the RadioAstron satellite: the
    orbit, the frequency coverage and the sensitivity. See
    \citet{tsuboi_09} and \citet{rauh} for more
    details. }\label{tab:ag}
\begin{tabular}{|l|r||r|r|}
\hline 
\multicolumn{2}{|c|}{ASTRO-G}&\multicolumn{2}{|c|}{RadioAstron}\\
\hline
\multicolumn{4}{|c|}{Orbital parameters}\\
\hline 
Apogee height   & 25,000 km&Apogee height   & 350,000 km\\
Perigee height   & 1,000 km  & Perigee height   &   25,000 km \\
Inclination angle& 31$^{\circ}$&Inclination angle&  51$^\circ$\\
Orbital period    & 7.5 hr       & Orbital period     & 8.5 days\\
\hline
\multicolumn{4}{|c|}{Observing frequency}\\
\hline
X-band &8.0 -- 8.8 GHz&P-band& 0.320 -- 0.328 GHz\\
K-band &20.6 -- 22.6 GHz&L-band& 1.64 -- 1.69 GHz\\
Q-band &41.0 -- 45.0 GHz&C-band& 4.80 -- 4.86 GHz\\
                                      &&K-band& 18.4 -- 25.1 GHz\\
\hline
\multicolumn{4}{|c|}{Antenna sensitivity (SEFD) [Jy]}\\
\hline
X-band &6100&P-band&15400\\
K-band &3600&L-band&2300\\
Q-band &7550&C-band&4400\\
              &       &K-band&6500\\
\hline
\end{tabular}\end{center}\end{table}

\subsection{An Analytical Study on the Performance of SFPR}\label{sec:m_analytic} 

A comprehensive analytical study on the propagation of
errors in the contributions listed in Equation 1 and in the
interpolation processes
into SFPR analysis for ground VLBI observations was presented in RD11.
Here, we expand this study to include the case of S-VLBI
observations.  The formulae listed in Table \ref{tab:eq} are a
slightly modified version of those in RD11, to account for the
atmospheric-free orbiting antenna in space-ground
baselines (i.e. a factor $\sqrt 2$ removed).  
We present the phase residual estimates for four frequency pairs, 
multiple frequency switching cycles, and typical model errors. 
The pairs of frequencies were selected to match the capabilities 
of RadioAstron (4.8/19 GHz) and ASTRO-G (8.4/22 GHz; 22/43 GHz),
and explore higher frequency S-VLBI (43/86 GHz). 
The frequency switching cycles were selected to match the frequency
agility of ASTRO-G and VLBA (1 minute) and to explore other regimes (0
minute, 2 minutes).
 
For comparison we include similar estimates for PR.

\begin{landscape}
\begin{table}
\caption{Formulae for estimating the residual phase errors for a ground-space 
baseline 
caused by propagation of errors in the corresponding models, given as
subscripts, during the analysis using SFPR techniques, at
$\nu^{high}$.  The subscripts ``A'' and ``B'' stand for the target and
calibrator sources, respectively; $\Delta \theta$ stands for the
source separation (or switching angle) and $T_{\rm swt}$ is the source
switching cycle.  The superscripts ``high'' and ``low'' stand for the
two frequencies $\nu^{high}$ and $\nu^{low}$ observed, respectively,
$T^{\nu}_{\rm swt}$ is the frequency switching cycle,
$R$ the frequency ratio (${\nu^{high} \over \nu^{low}}$), and
$\theta$ stands for the magnitude of the core shift between the two
observed frequencies.  
$D$ is the baseline length, $\Delta P$ represents the combined contribution
of Earth Orientation Parameters and antenna position errors, 
and $\Delta s^c$ is the error in
the position of the calibrator source. $C_w$ describes the weather
conditions, with values equal to 1, 2 and 4 for good, typical, and
poor tropospheric conditions, respectively. $\Delta l_z$ stands for
the tropospheric systematic zenith excess path error, and $\Delta I_v$
is the ionospheric vertical TEC systematic error. $Z_g$, $Z_i$ and
$Z_F$ are the zenith angles which describe the elevation dependence of
the atmospheric line-of-sight excess path at different altitudes - for
a detailed explanation of these parameters see A07, and RD11.}\label{tab:eq}
\begin{tabular}{ll}
\hline
Error Contributions & SFPR residual phase error  \\ \hline
{\it Atmospheric Models:} & \\
Dynamic Troposphere & 
$ 
\sigma \phi^{{\rm high}}_{{\rm dtrp}} [{\rm deg}] \approx
R \, 27 \, C_w \, \left( {{\rm \nu^{low}}[{\rm GHz}] \over 43 {\rm
    GHz}} \right) \left( {\sec{Z_g} \over \sec{45^o}} \right)^{1/2}
\times \left( {{\rm T^{\nu}_{\rm swt}} [{\rm s}] \over {\rm 60s}} \right)^{5/6},
$ \\
Static Troposphere &
$ 
\sigma \phi^{{\rm high}}_{{\rm strp}} [{\rm deg}] \approx  R
\, 76 \left({{\rm \nu^{\ low}} [{\rm GHz}] \over 43 {\rm GHz}}\right)
\left({\Delta l_z [{\rm cm}] \over 3 {\rm cm}}\right) \left({\theta_A
  [^0] \over 2^o}\right) \left({\cos{Z_g} \over
  \cos{45^o}}\right)^{-1} \left({\tan{Z_g} \over \tan{45^o}\ }\right)
\approx 0 ,
$ \\
Dynamic Ionosphere &
$
\sigma \phi^{{\rm high}}_{{\rm dion}} [{\rm deg}] \approx
\left(R-1/R\right) \, 0.46 \ \, \left({\sec{Z_i} \over
  \sec{43^o}}\right)^{1/2} \left({\nu^{{\rm low}} [{\rm GHz}] \over 43
  {\rm GHz}}\right)^{-1} 
    \times \left[0.21 \left({\rm
    T_{\rm swt} [s] \over 60 s}\right) + \left({\sec{Z_i} \over
    \sec{43^o}}\right) \left({\Delta \theta [^o] \over
    2^o}\right)\right]^{\ 5/6} ,
$ \\
Static Ionosphere &
$
\sigma\phi^{{\rm high}}_{{\rm sion}} [{\rm deg}] \approx 
\left(R-1/R\right)\, 2.7 \, \left({\nu^{{\rm low}} [{\rm GHz}] \over
  43 {\rm GHz}}\right)^{-1} \left({\Delta I_V [{\rm TECU}] \over 6
  {\rm TECU}}\right) \left({\Delta \theta [{\rm deg}] \over
  2^o}\right) 
  \times \left({\cos{Z_F} \over
  \cos{41^o}}\right)^{-1} \left({\tan{Z_F} \over \tan{41^o}}\right )
,
$ \\
{\it Geometric Models:} & \\
Source Position & 
$
\sigma \phi^{{\rm high}}_{{\rm \Delta s}} [{\rm deg}] \approx R \, 16
\left({\nu^{low} [{\rm GHz}] \over 43 {\rm GHz}}\right)
\left({\rm D [km] \over 6000 km}\right) \left({\Delta s^c [{\rm mas}] 
       \over 0.3 {\rm mas}}\right)
\times \left({\theta_{\rm A} [{\rm deg}] \over 2^o}\right)
\approx 0  ,
$ \\
Telescope Position& 
$
\sigma \phi^{{\rm high}}_{{\rm bl}} [{\rm deg}] \approx R \, 18
\left({\nu^{{\rm low}} [{\rm GHz}] \over 43 {\rm GHz}}\right)
\left({\rm \Delta P [cm] \over 1 cm}\right) 
\times \left({\rm \theta_A [deg] \over 2^o}\right)
\approx 0, 
$ \\
{\it Others:} & \\ 
Instrumental Contribution & 
$
\sigma \phi^{{\rm high}}_{{\rm inst}} = \left(\phi_{{\rm A,inst}}^{{\rm high}} - R
\,.\, \tilde \phi_{{\rm \ A,inst}}^{{\rm low}}\right) -
\left(\phi_{{\rm B,inst}}^{{\rm high}} - R \,.\, \tilde \phi_{{\rm
    B,inst}}^{{\rm low}}\right) \approx 0,
$ \\
Thermal Noise & $\sigma \phi^{{\rm high}}_{{\rm thermal}} = \sqrt
{ (\sigma \phi^{\nu^{{\rm low}}, \nu^{{\rm high}}}_{{\rm
      A,thermal}})^2 + (\sigma \phi^{\nu^{{\rm low}}, \nu^{{\rm
        high}}}_{{\rm B,thermal}})^2 + (\sigma \phi^{\nu{{\rm
        high}}}_{{\rm A,B,thermal}})^2}$ \\
 \hline
\end{tabular}
\end{table}
\end{landscape}

\subsection{A Simulation Study on the Performance of SFPR}\label{sec:m_sim}

It has been shown \citep{pradel_06,asaki_07,honma_08} that simple
analytical analysis is insufficient for the estimation of astrometric
errors in VLBI observations.
We have carried out simulation studies to describe the performance of
SFPR techniques with S-VLBI observations, both for astrometry and
increased sensitivity purposes.  The procedure consists of generating
synthetic S-VLBI datasets using the ARIS (A07) simulation tool for a
given observing configuration and typical values for model errors, as
listed in Table \ref{tab:par}, and carrying out the SFPR data analysis
with AIPS.  A detailed description of the models implemented in ARIS
can be found in A07; the SFPR analysis with AIPS is described in RD11
and \citet{dodson_m31}.  The outcome of each iteration is a SFPR-map.
The figures of merit used to characterize the performance 
for a set of parameters in ARIS
are: the average fractional peak flux recovery, which is the ratio between the 
map peak and the model source fluxes, and the astrometric error, which is 
the offset of the peak from the center of the map. 
Larger values for flux recovery and smaller
values for astrometric error quantities are indicative of a better performance.
The results for a given observing configuration comprise of multiple
simulation cycles, each with a changing relative orientation of the
source pair along the four cardinal directions.  Each cycle was done
with independently generated random values for the tropospheric,
ionospheric and geometric parameters.

The results from our SFPR simulation studies are presented in Section
3.  Furthermore, the results from these simulations can be
extrapolated to other regions of the parameter space, as discussed in
Section 4.

\begin{landscape}
\begin{table}
\caption{Region of the Parameter Space explored in our Simulation Studies}\label{tab:par}
\begin{tabular}{ll}
\hline
Parameter Description  & Parameter Values \\
\hline
Space segment & 1 satellite antenna with ASTRO-G-like orbit and diameter  \\
Ground segment & a) 10 antennas (the VLBA array) \\
               & b) 6 antennas (sparse non-uniform array, 
with simultaneous dual frequency capability) \\
Observing Frequencies & $\nu^{low}=22\,$GHz (K-band);
$\nu^{high}=43\,$GHz  (Q-band)\\
Source observing configuration & Source switching angle ($\Delta
\theta$) : 0.1, 0.5, 1, 2, 3, 4, 5 degrees \\  
                               & Source switching cycle ($T_{\rm swt}$): 
3,4,6,8,10 minutes \\
Source Model & Compact structure, S=1 Jansky \\
Frequency observing configuration & a) Fast frequency switching 
(T$^{\nu}_{\rm swt}=1$ minute) \\
                                   & b) Simultaneous dual frequency observations (T$^{\nu}_{\rm swt}=0$ sec) \\
Weather conditions & a) Good ($C_w=1$) \\
                   & b) Typical ($C_w=2$) \\
                   & c) Poor ($C_w=4$) \\
Model Errors$^1$ & a) ODDA: 2, 4, 6, 8, 16, 32, 64 and 128 cm \\
             & b) `Typical' values for propagation medium: \\ 
             &    ($\Delta l_z = 3$cm; $\Delta I_v=6$TECU) \\
             & c) Ground antenna position errors, 1 cm \\
Studies & a) Astrometry \\
        & b) Phase Coherence \\
\hline
\end{tabular}

\vspace*{1cm}

{\footnotesize {\bf (1):} {We have used typical values for the parameter model errors, as listed in A07}} 

\end{table}
\end{landscape}

\section{RESULTS}\label{sec:results}

\subsection{Analytical Study of SFPR for S-VLBI} \label{sec:r_analytic}

\begin{figure}
\caption{Residual phase error budgets for a S-VLBI
  baseline for: (a) PR observations at 8.4, 22 and 43 GHz; (b) and (c) SFPR 
  observations at 4.8/19 GHz, 
  8.4/22 GHz, 22/43 GHz, and 43/86 GHz, estimated with our analytical studies.  The
  plots show the individual contributions arising from typical errors in the 
  geometrical, both for the antenna/orbit and reference source 
  coordinates (Geo), tropospheric (Trop) and ionospheric (Iono)
  models. The values have been estimated using the formulae in Table
  \ref{tab:eq} at $\nu^{high}$, for SFPR, and in A07, for PR. 
  The PR estimates are for alternating observations of a pair of
  sources 2 degrees apart, with a switching cycle of 1 minute. 
 For SFPR, we show estimates of two configurations: 
 b) solid bars are for alternating observations between two frequencies, with a switching cycle $T^{\nu}_{\rm swt}$ 
 of 1 minute, of a pair of sources 2 degrees away, with a source switching cycle $T_{\rm swt}$ of
 4 minutes;  open bars show the same, for a frequency switching
 cycle of 2 minutes. c) is the same as b), with simultaneous dual frequency
 observations, i.e. T$^{\nu}_{\rm swt}=0$.
  Other relevant parameters which are kept
  common for all estimates are, satellite ODDA 10 cm, typical
  tropospheric and ionospheric parameter errors ($\Delta l_z\,=\,3\,$cm, $\Delta I_v\,=\,6\,$TECU) 
  and good weather conditions.
  Note the different scales in the vertical axis. The horizontal
  dashed line in all figures corresponds to the 
  largest tropospheric contribution at 43 GHz in a).
}\label{fig1}
\centering
\includegraphics[width=7.8cm]{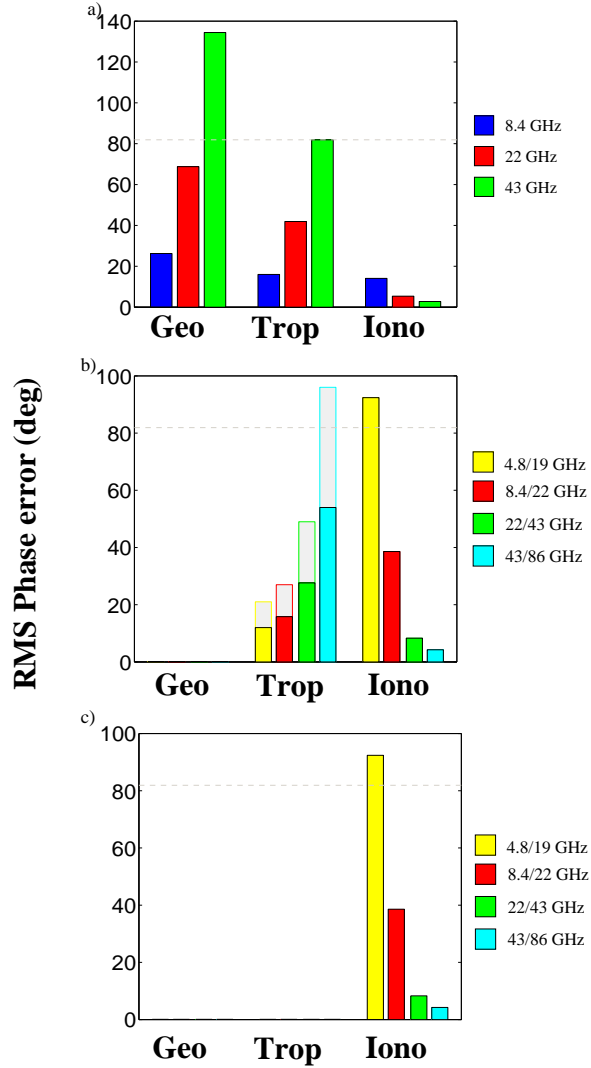} 
\end{figure}

Figure \ref{fig1} shows comparative RMS residual error budgets
estimated for PR (Figure \ref{fig1}a) and SFPR (Figure \ref{fig1}b and
Figure \ref{fig1}c) applied to S-VLBI
observations.  The error budgets comprise contributions arising
from inaccuracies in the geometric (i.e. satellite orbit and reference
source coordinates errors), tropospheric and ionospheric models;
these are the dominant sources of errors in a non Signal to Noise
Ratio (SNR) limited case.
The case of noise dominated observations will be addressed in a future
study. The observing configurations are: Figure \ref{fig1}a) for conventional
PR observations at 8.4, 22 and 43 GHz with a source switching cycle
of 1 minute;
Figure \ref{fig1}b) for SFPR observations at four pairs of
frequencies ($\nu^{low} / \nu^{high}$), namely, 4.8/19, 8.4/22, 22/43 and
43/86 GHz, with frequency switching cycles T$^{\nu}_{\rm swt}$ of 1 and 2
minutes, and source switching cycle T$_{\rm swt}$ of 4 minutes, and
Figure \ref{fig1}c) is the same as Figure \ref{fig1}b) but with
simultaneous dual frequency observations, i.e.  T$^{\nu}_{\rm swt}$=0.
In all cases a source pair angular separation of 2 degrees,
a satellite orbit height at apogee of 25,000 km, an orbit error ODDA equal
to 10 cm, `good' weather conditions, and typical values for the
remaining model errors have been used.  
The estimated values for SFPR technique have been
derived using the formulae in Table \ref{tab:eq}, and those for
conventional PR with the formulae in A07, for a ground-space baseline.

We briefly describe the relative strengths from the individual
error contributions in Figure \ref{fig1} as a function of the observing
frequency for each technique.
The orbit error is responsible for the largest residual phase
contribution using PR techniques, at all frequencies. This is the case
for any realistically achievable orbit error, as discussed in A07.  The
tropospheric errors constitute the next largest contribution, with
residuals that are linearly proportional to the observing
frequency. Instead, the ionospheric residual contribution is much
reduced at 22 and 43 GHz, with respect to that at 8.4 GHz.
In the SFPR analysis the effect from the orbit errors is
completely compensated, as shown in Figures \ref{fig1}b,c. 
Using a frequency switching cycle of 
1 minute (shown with a solid color bar) the tropospheric (i.e. dynamic) contribution, 
arising from interpolation errors to the times of the higher 
frequency observations,
is the largest in observations with $\nu^{low} \ge 22$ GHz.
This contribution is significantly
increased with a frequency switching cycle of 2 minutes 
(shown with a light gray bar with a color coded outline).  
At the lower frequency pairs (8.4 to 22 GHz, 4.8 to 19 GHz bands)
the ionospheric contribution is dominant. This contribution can be reduced by
having a closer pair of sources, while it is independent of the frequency 
switching cycle. 
Note that the residual tropospheric contribution in 
SFPR is smaller than in PR because the
static contribution is fully compensated using
same line-of-sight observations at the two 
frequencies. 
For the case of simultaneous
dual frequency SFPR observations (i.e. $T^{\nu}_{\rm swt}$=0) the dynamic tropospheric 
errors are also fully compensated, leaving only the much smaller (with $\nu^{low} \ge 22$ GHz) 
ionospheric contribution as shown in \ref{fig1}c. 
These are typically less than 10\% of those from the satellite orbit and
tropospheric errors in PR at the highest S-VLBI frequency bands. 
It is worth mentioning that the analytical studies predict
that variations in the source switching cycle in 
SFPR observations have no significant impact in the RMS estimated 
residual phase budget; the same applies for variations in the source 
pair angular separation for the pairs with $\nu^{low} \ge 22$ GHz.
The reason being that these affect only the weak ionospheric residuals.
Based on these results we proceeded to do a more complete
simulation-based investigation.

\subsection{Simulation Studies for Astrometry with Space VLBI}\label{sec:r_astrometry} 

For our simulation studies we have closely followed the style of the PR
feasibility studies for S-VLBI reported in A07, here applied to SFPR. 
The major effects we wished to explore are the effects of 
i) frequency switching cycle, ii) weather, iii) source pair angular separation,
iv) source switching cycle and v) satellite orbit errors. 
The frequency switching cycle $T_{\rm swt}^{\nu} = 1$ minute, and the observing frequencies
($\nu^{low} = 22$, $\nu^{high}= 43$ GHz), are compatible with the ASTRO-G mission specifications as well as 
the VLBA. 
Full sets of solutions were generated for the whole 2-dimensional grid
of source pair switching angles ($\Delta \theta=$ 0.1, 0.5, 1, 2, 3, 4,
5$^\circ$) and source switching cycles ($T_{\rm swt}=$3, 4, 6, 8 and 10 minutes),
for fast frequency switching SFPR observations, between 22 and 43 GHz,
with $T_{\rm swt}^{\nu}= 1$ minute, and for simultaneous dual frequency
observations, $T_{\rm swt}^{\nu}= 0$.  
These were repeated for good, typical and poor weather
conditions -- as defined in A07. The  ODDA is 8 cm in all cases, 
except where otherwise stated. Other parameters were as in A07.

\begin{figure}
\epsscale{1}
\plotone{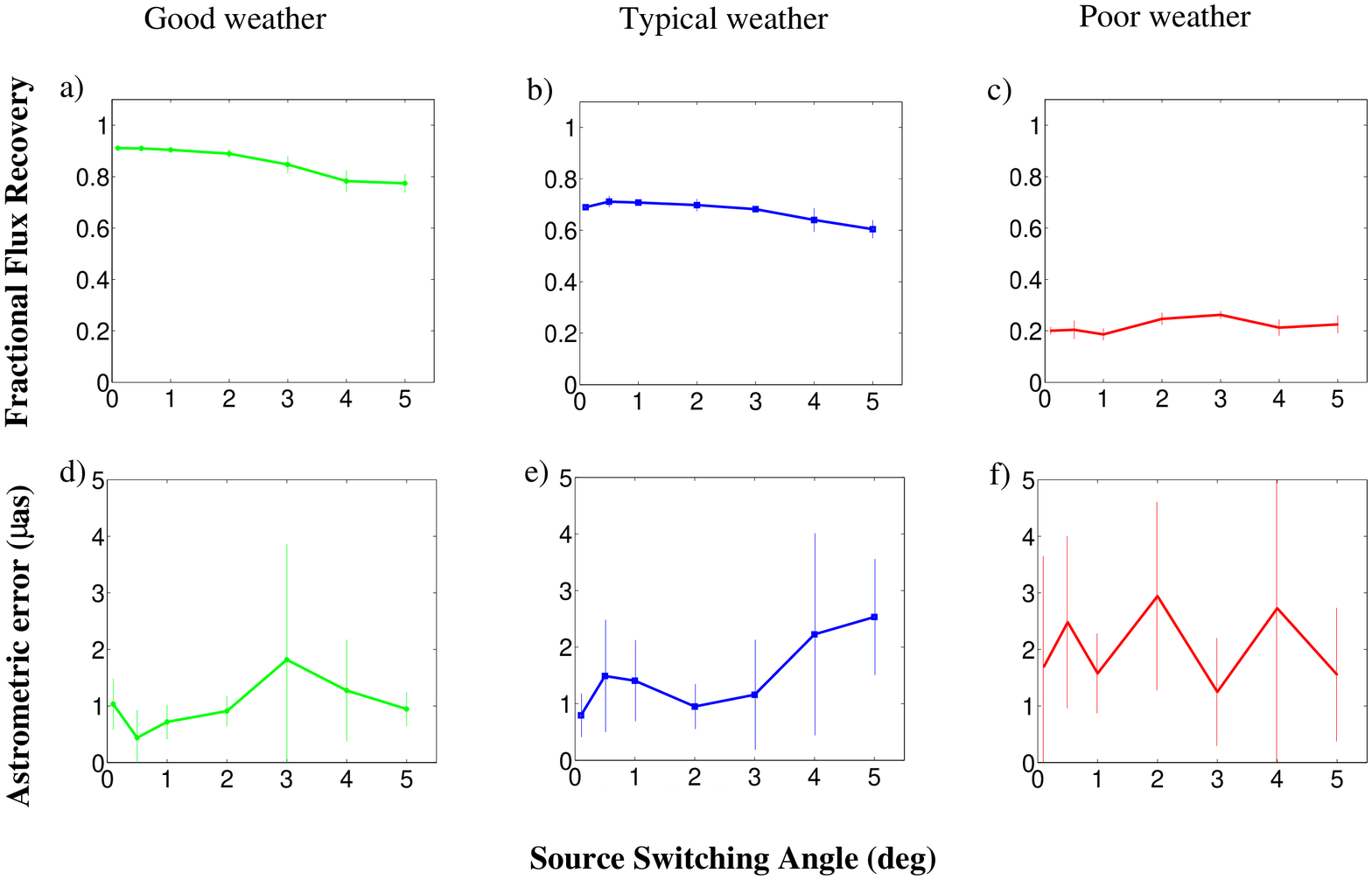}
\caption{Fractional peak flux recovery (top row: a--c) and 
astrometric error (bottom row: d--f) quantities as a function of
source separation, measured from the SFPR maps at 43 GHz.
The results plotted correspond to the mean values of 4 simulations; the
error bars are the $\pm$RMS values. Simulations were performed
with a frequency switching cycle $T^{\nu}_{\rm swt}$ of 1 minute, a source switching
cycle $T_{\rm swt}$ of 6 minutes,  ODDA orbit errors of 8 cm and
source separations $\Delta \theta$ of 0.1, 0.5, 1, 2, 3, 4 and 5 degrees.}
\label{fig2}
\end{figure}

Here we present a subset of these 
simulations, as the figures of merit were found to have a flat distribution.
For the sake of clarity we present two 1-dimensional cuts which fully
convey the outcomes.  Figure \ref{fig2} presents the peak flux recovery and
the astrometric accuracy quantities through the dataset against all
source switching angles, for a source switching cycle of 6
minutes. These results are presented for good (Figure 2a,d), typical
(Figure 2b,e) and poor (Figure 2c,f) weather conditions.  The figures
of merit show no significant variation across this range, with mean
values of 86\%, 67\% and 22\% and 1, 1.5 and 2 $\mu$as for good,
typical and poor weather, respectively.

\begin{figure}
\epsscale{1}
\plotone{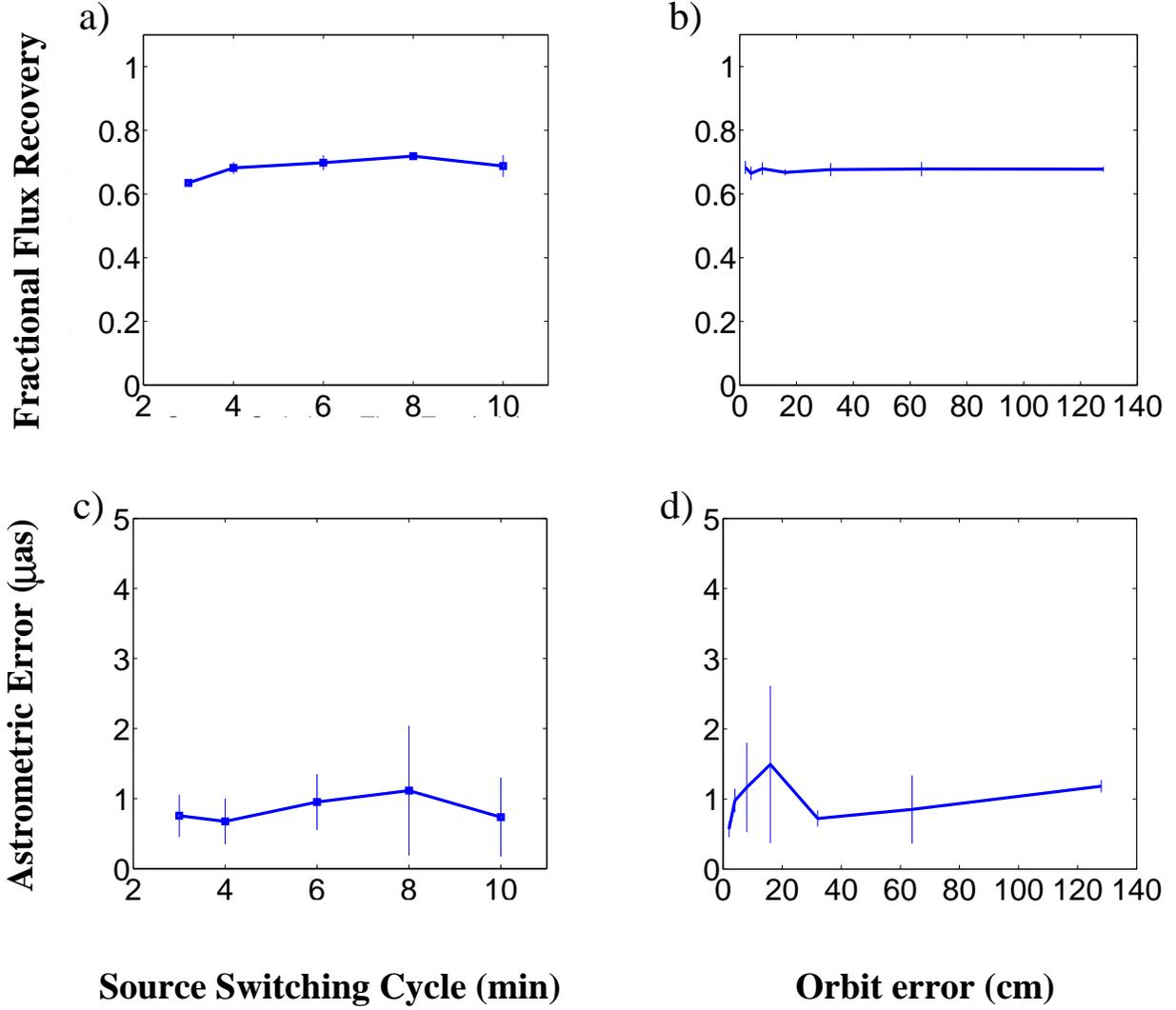}
\caption{Fractional peak flux recovery (top row: a, b) and 
astrometric error (bottom row: c, d) quantities as a function of source switching
cycle (left: a, c) and ODDA orbit error (right: b, d), measured from the SFPR maps
at 43 GHz. 
The results plotted correspond to the mean values of 4 simulations;
the error bars are the $\pm$RMS values. Simulations were performed under
typical weather conditions, with a frequency switching cycle
$T^{\nu}_{\rm swt}$ of 1 minute. On the left a source separation
$\Delta \theta$ of 2$^\circ$, orbit error of 8 cm, and source switching cycles $T_{\rm swt}$ of
3, 4, 6, 8 and 10 minutes. On the right, a source separation
$\Delta \theta$ of 3$^\circ$, a source switching cycle of 8 minutes and 
ODDA orbit errors from 2 to 128 cm, doubling between simulations.  
}
\label{fig3}
\end{figure}

Figure \ref{fig3}a,c) show the one-dimensional cuts through the dataset
against all source switching cycles, for a switching angle of 2$^\circ$
and typical weather conditions, showing the fractional peak flux recovery and
astrometric error quantities, respectively.  Here too, no significant
variation is found across the tested range with mean values of 68\%,
and 0.9\,$\mu$as.

Of special relevance for S-VLBI is the propagation of errors in
the orbit determination into the astrometric analysis.  Figure
\ref{fig3}b,d show the figures of merit
obtained for ODDA values equal to 2, 4, 8, 16, 32, 64, 128 cm, for SFPR
observations of a pair of sources 3 degrees away, using a source
switching cycle of 8 minutes, with typical weather conditions. As in
previous cases, no significant changes in the values of the figures of
merit were seen across the tested range, with mean values of 
68\% and 1\,$\mu$as.

\begin{figure}
\epsscale{1}
\plotone{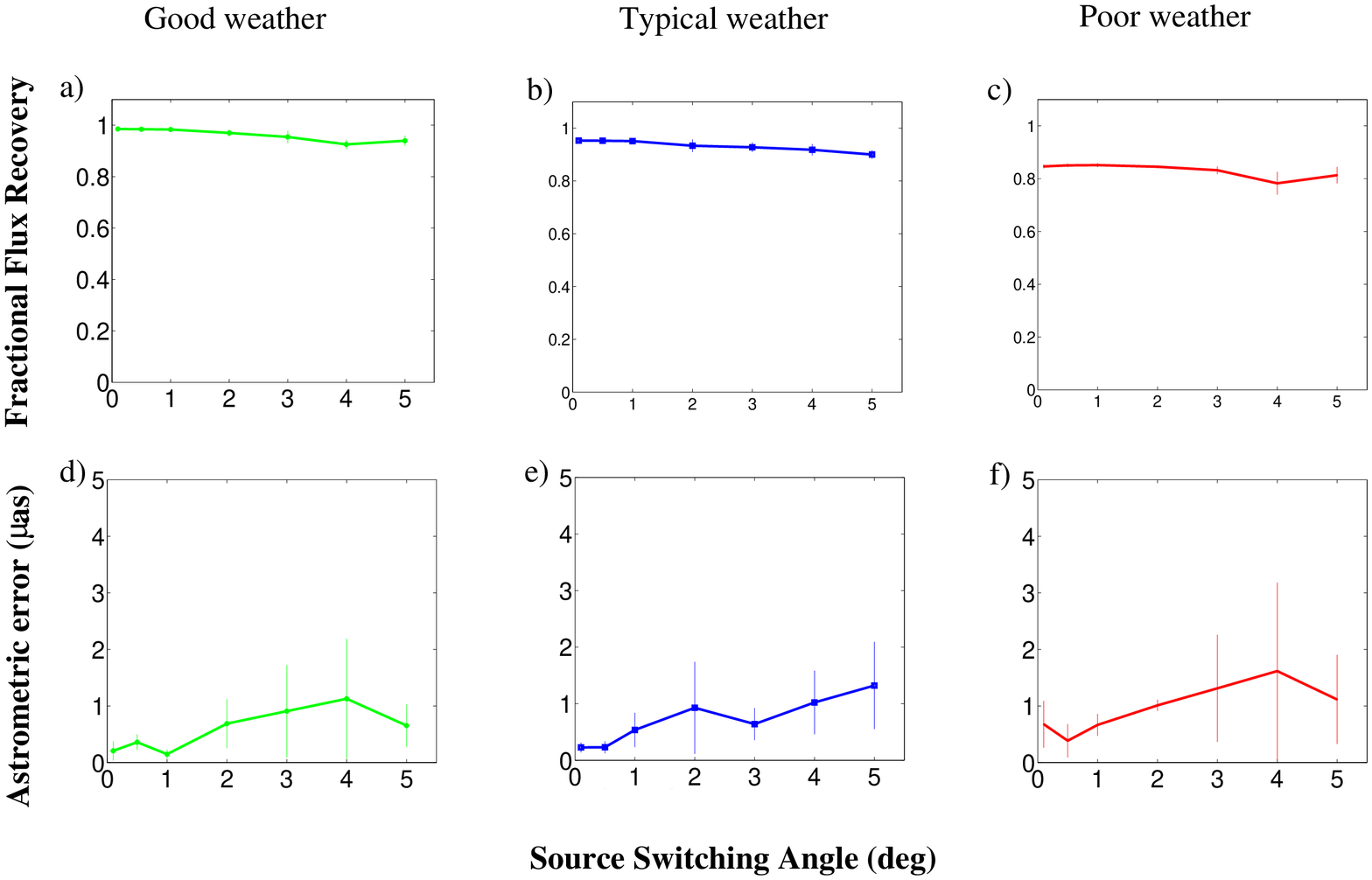}
\caption{Same as Figure 2, but for simulations using simultaneous dual frequency
observations, $T^{\nu}_{\rm swt}=0$.}
\label{fig4}
\end{figure}

The benefits of carrying out simultaneous dual frequency SFPR observations 
has been recognized from our previous studies (RD11). 
Hence, we have included this configuration in our
simulations, 
even though this is not a capability of the ASTRO-G mission nor the VLBA
ground array. Thereby we are able to characterize
the benefits in comparison with fast frequency switching observations
between 22 and 43 GHz. The rest of the parameters for the satellite and ground
array antennas were kept the same as in previous simulations.
Figure \ref{fig4} is equivalent to Figure \ref{fig2}, but for 
simultaneous dual frequency SFPR observations at 22 and 43 GHz.
The simulations shown in Figure \ref{fig4} were carried out using the 
VLBA as the ground array, for consistency. In this case the 
flux recovery and astrometric error mean values are 
96\%, 93\% and 83\% and 0.6, 0.7 and 1 $\mu$as for good,
typical and poor weather, respectively.
We also repeated the simulations and confirmed that the results are
not significantly different for the case when the VLBA is replaced with a more
realistic ground array. This array consisted of antennas that either have, or
have expressed plans for, a suitable simultaneous 22/43 GHz receiving
system, for example arrays with quasi-optics systems such as the Korean
VLBI Network (KVN) \citep{kvn_ref}. 
This realistic ground array was comprised of 6 antennas:
3 KVN antennas (Korea), Yebes-40m (Spain), Effelsberg (Germany) and
Shanghai-65m (China). 

\begin{figure}
\centering
\includegraphics[width=10cm]{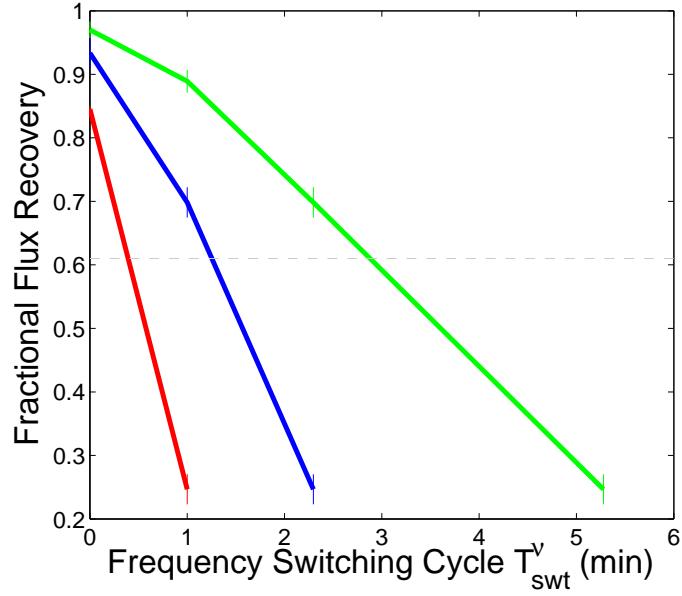} 
\caption{Performance of SFPR observations 
at 22/43 GHz for a range of frequency switching cycles, under
good (right-most line), typical (middle line) and poor (left-most line) 
weather conditions. 
In the online version these lines are shown in green, blue and red, respectively.
The values in the plot have been derived
directly from the simulations, for $T_{\rm swt}^{\nu}$ = 0, 1 minutes,
or indirectly, 
scaling those cycle times by the tropospheric $C_w$ parameter raised to the power of 6/5. 
All simulations have been carried out using $\Delta\theta$=2$^\circ$, 
ODDA=8 cm and T$_{\rm swt}$=6 minutes.
The rapid fall in the fractional peak flux recovered for poor weather
indicates that a frequency switching cycle faster than 0.4
minutes is required for `reliable' observations. Under typical
weather conditions $T^{\nu}_{\rm swt} \le $ 1.3 minutes are required, 
while for good weather $T^{\nu}_{\rm swt} \sim $ 2.9 minutes would be acceptable. 
Based on these results we recommend fast frequency switching cycles of $\sim$ 1 minute
for good and typical weathers, and simultaneous dual frequency
observations under poor
weather conditions, at $\nu^{high}=43$ GHz.
}
\label{fig5}
\end{figure}

The choices for T$^\nu_{\rm swt}$ in the simulations was driven by the
specifications of the various interferomic arrays discussed. The
VLBA/ASTRO-G minimum switching time is 1 minute and for KVN it is 0.
Nevertheless, given that the frequency switching
time, T$^\nu_{\rm swt}$, and the weather scale factor, $C_w$, 
affect only the dynamic troposphere error contribution, we can combine 
our simulation results to describe the performance of SFPR observations 
using frequency switching cycles in the range ca. 0 to 5 minutes,
under all weather conditions.
Figure \ref{fig5} shows 
that, at $\nu^{\rm high}$=43 GHz, frequency switching cycles faster
than 0.4, 1.3 and 2.9 minutes are required to maintain fractional peak
flux recovery greater than 61\%, equivalent to a RMS phase error of 1
radian, for poor, typical and good weather conditions, respectively.

\subsection{Simulation Studies for Increased Coherence Time in Space VLBI}\label{sec:r_coh} 

We present here the results of our simulation studies to test the
feasibility of dual frequency observations to increase the sensitivity
of S-VLBI, by increasing the coherence time.  
We have carried out comparative FPT-calibration simulation studies using fast frequency 
switching observations with a 1 minute cycle, and simultaneous
dual frequency observations, at 22 and 43 GHz, of a single source as described in Section 2.1. 
Also, for comparison, we carried out simulations using single frequency observations 
at 43 GHz, followed by self calibration analysis. 
We used an ODDA orbit error of 8 cm and typical values for the rest of
the parameter model errors.
All the simulations were carried out for good, typical and poor weather conditions.
In all cases, the analysis in {\sc AIPS} was repeated multiple times using different
temporal solution intervals in the task {\sc CALIB}, from 2 seconds to $\sim$8 hours
(in steps doubling the interval) prior to the Fourier inversion
to generate the image at 43 GHz. 
In each case, the fractional peak flux recovery was measured from the image generated using only
ground-space baselines.

\begin{figure}
\centering
\includegraphics[width=7cm]{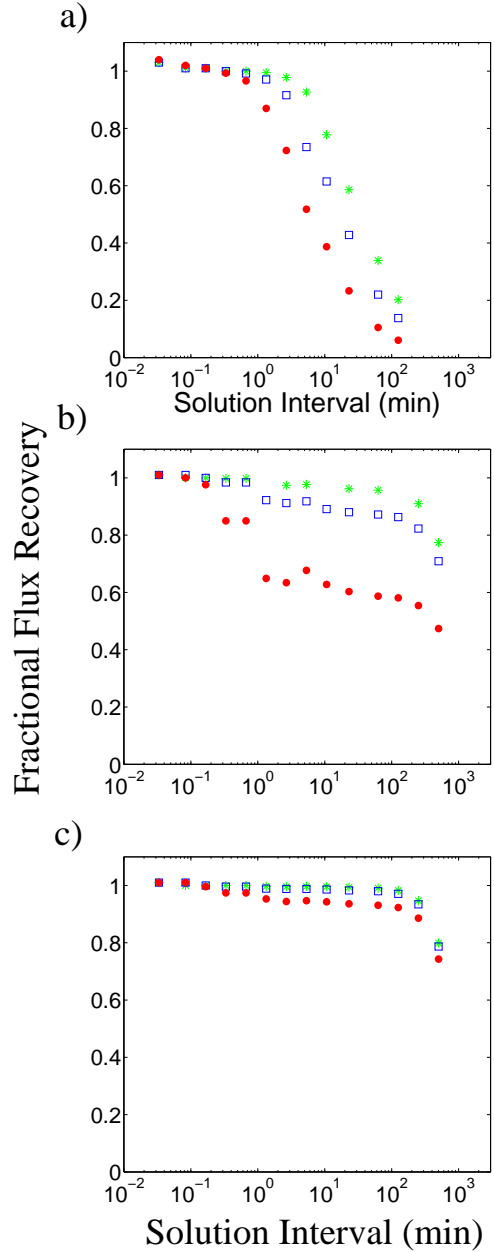} 
\caption{ The variation of fractional peak flux recovered as measured from
  images at 43 GHz as a function of the temporal solution interval in the AIPS
  task {\sc CALIB} prior to Fourier inversion of the simulated data.  The
  images have been generated with data from the ground-space baselines
  only, using different calibration schemes: a) self calibration only;
  b) FPT-calibration for dual frequency observations at 22 and 43 GHz
  with a fast frequency switching cycle $T^{\nu}_{\rm swt}$ of 1 minute,
  followed by self calibration; and, c) same as b), but with
  simultaneous dual frequency observations, $T^{\nu}_{\rm swt}=0$.  
  All simulations are carried out with an ODDA orbit error of 8 cm, 
  and typical values for the other of model parameter errors.   For
  each case, the three weather conditions are shown: good (stars),
  typical (squares), and bad (circles).  In the online version those
  are shown in green, blue and red, respectively.
}
\label{fig6}
\end{figure}

Figure \ref{fig6}a
shows the measured fractional peak flux recovery 
plotted against the temporal solution interval for the self-calibration analysis case, 
for all weathers. Taking the usual measure for
coherence time as the point where the flux recovery falls below 61\%
we estimate coherence times for the space to ground baselines of 4, 10 and
20 minutes for poor, typical and good weather respectively. 

However, if the 
tropospheric fluctuations, and orbital errors, are compensated using the observations 
at 22 GHz the coherence time is expected to increase.
Figure \ref{fig6}b
corresponds to the case of using dual frequency
observations with a fast frequency switching cycle of 1 minute, which shows
coherence times for good and typical weathers extending to many hours
(across the whole simulation run)-- which would lead to an order of
magnitude improvement of the minimum flux for a detectable source. We
find that, unsurprisingly, the poor weather case quickly falls (in 0.5 hours) to a
plateau of fractional peak flux recovery around 50--60\%, which would
cast doubt as to whether this approach would work in poor
weather. 
Finally, the results from simulations using simultaneous dual
frequency observations are shown in Figure \ref{fig6}c.
In this case, a coherence time up to many hours is estimated under all
weather conditions. 
Note that the remaining long timescale  ionospheric and instrumental
(i.e. dispersive) residuals using only dual
frequency FPT calibration prevents one from achieving astrometry.
Also, these are responsible of the decrease in flux observed in Figs
\ref{fig6}b and \ref{fig6}c at the longest temporal solution intervals.
Using SFPR observations, which include a second source, the flux
curve remains flat and also would enable astrometry.

\section{Discussion}\label{sec:discussion}

\subsection{Suitability of SFPR for astrometry with S-VLBI} 

The SFPR technique addresses the issues that limit the application of 
PR, which works well for ground arrays, 
to S-VLBI and is a 
feasible technique to achieve astrometry and enhanced sensitivity 
even at the highest frequencies.
The use of conventional PR calibration techniques for S-VLBI
(especially at frequencies $> 22$GHz) poses challenges arising
from the satellite antenna position errors, which are much larger than for Earth
based antennas, and poor sensitivity due to the small orbiting
antennas which, in turn, result in the shortage of nearby suitable
calibrator sources, plus constraints in satellite operations,
i.e. for fast source switching, required to compensate fast tropospheric
fluctuations. 
The SFPR technique compensates for tropospheric and geometric non-dispersive 
model errors using observations at two frequencies, 
either with fast frequency switching or simultaneously observed. 
We have carried out analytical and simulation studies to verify the
feasibility 
of SFPR applied to S-VLBI, both with positive outcomes.  
Our simulation studies comprise SFPR observations of a satellite antenna with a
network of ground telescopes, at 22 and 43 GHz, of compact sources  
with a range of angular separations between 0.1 to 5
degrees, source switching cycles between 3 and 10 minutes, 
different weather conditions, and satellite 
orbit determination discrepancies at apogee (ODDA) ranging from 2 to 128 cm. 
The simulations are based on the ASTRO-G mission specifications 
for the satellite antenna, which is capable of frequency switching
cycles of 1 minute, except for the subset of simulations with
simultaneous dual frequency observations.  The ground array was the VLBA, except for the case of
simulations with a `realistic' ground array, which was limited to 
those antennas which support simultaneous 
dual frequency observations or have plans towards this capability. 
The analytical studies comprise a wider range of frequencies, starting at 5 GHz 
to match RadioAstron specifications, and up to 86 GHz.

We use 
two figures of merit to characterize the performance from the simulations. 
The fractional peak flux recovery quantity shows an improved performance 
with shorter frequency switching cycles, especially under poor weather 
conditions, and for a given cycle, it deteriorates with worsening 
weather; in comparison it shows little dependence on the pair angular
separation or the source switching cycle.
The analytical results show that this tendency continues to apply 
in the tropospheric dominated regime ($\nu^{low}\ge 22$ GHz).
Instead, when $\nu^{low}$ is much less than this, the ionospheric errors
dominate and increase with pair angular separation. 
In all cases the performance is independent of the orbit errors.

The results from our simulations show 
that `reliable' (taken as flux recovery over 61\%) 
astrometric measurements between 22 and 43 GHz
at the $\mu$-arcsecond precision level are achievable with 
SFPR observations 
with frequency switching cycles $T^\nu_{\rm swt} <$ 0.4, 1.3 and 2.9 minutes under poor, typical
and good weather conditions, respectively. 
The astrometric error quantity in our simulations does not vary
significantly under the tested conditions, 
irrespective of the orbit determination error,
the angular separation between the two sources $\Delta \theta$, and the source switching
cycle $T_{\rm swt}$. 
This behavior can be easily
explained by the two-step strategy of SFPR to compensate
errors of different nature.  
The dynamic tropospheric residuals, which are the dominant contribution
if $\nu^{low} \ge 15-22 GHz$, are proportional
to the frequency switching cycle, as shown in Table \ref{tab:eq}.
The random nature of the residual errors is expected to cause blurring
effects in the final SFPR-image, 
i.e. while the scattering of the phases is expected to  
decrease the peak flux in the image, 
the peak will not be shifted from the origin.
Only the remaining non-dispersive effects are sensitive to $\Delta \theta$ and $T_{\rm swt}$, and these 
are weak at the high frequencies of interest here.
Therefore, the {\sc sfpr} route holds great promise for S-VLBI since any orbit
errors (as well as ground antennae coordinate errors) are fully
removed in the analysis, therefore alleviating the constraint on orbit
determination accuracy. This is of importance as the ASTRO-G team has shown that one of the
greatest S-VLBI challenges is to accurately measure the antenna
position to a few centimeters for a $\sim$ 7-hour long period highly
elliptical orbit, with an apogee height of 25,000 km, hence any
astrometric approach which could circumnavigate that requirement would
be of major benefit.
Also, the conditions for a suitable {\sc sfpr}
calibrator source are much more relaxed than in conventional {\sc pr}, the angular separation 
between sources can be up
to several degrees, and the observing source duty cycle up to several 
minutes.

\subsection{Suitability for detection of weak sources} 
The SFPR technique offers a technical solution to alleviate the
sensitivity issue in S-VLBI observations that arises from the long
baselines and small satellite antennas, especially for observations at
high frequencies where the coherence time is limited by the rapid
tropospheric fluctuations.  The requirements for a suitable PR
calibrator are increasingly difficult to meet at increasing
frequencies in general, and all the more so for S-VLBI. 
Alternatively, the tropospheric compensation at the target observing frequency can be
achieved using detections of the same source at a lower frequency;
this results in increased sensitivity at the target frequency.
Our simulations characterize the performance
of dual frequency tropospheric calibration with S-VLBI
at 22 and 43 GHz, using both fast frequency switching (T$^\nu_{\rm swt}$=1min) and
simultaneous observations (T$^\nu_{\rm swt}$=0).  The results show that
increased coherence times up to several hours at 43 GHz (compared to
$\sim$ a few minutes of tropospheric coherence times) are achievable under
any weather conditions with simultaneous dual frequency
observations. More moderate benefits are obtained with fast frequency
switching observations.  Such an increase in coherence time is
equivalent to an increase in sensitivity by a factor of ten.
Therefore dual frequency observations enable increased
sensitivity for weak sources, one of the major issues for space
VLBI, especially at the high frequencies and with very large orbits like 
RadioAstron.  Achieving high sensitivity
S-VLBI observations is of paramount importance to address the key
science goals of S-VLBI missions.

\subsection{Benefits from using simultaneous dual frequency
  observations}
The residual phase error budget in SFPR observations using fast
frequency switching is dominated by tropospheric terms, for
$\nu^{low} \sim 22$ GHz and higher. 
For a given frequency switching cycle the tropospheric residual
estimates increase linearly with the target frequency $\nu^{high}$
 (see formulae in Table \ref{tab:eq}).
This is a result of the imperfect compensation of the rapid tropospheric 
fluctuations, which requires matching frequency switching times, 
especially at higher frequencies.
Our simulations show that a frequency switching cycle of 1 minute 
produces high precision astrometric estimates and increased coherence for a wide
range of observing configurations at 43 GHz under good and typical weather conditions,
but not for poor weather.  In this case
frequency switching cycles of 0.4 minutes or less are required. 
At higher frequencies, increasingly faster switching cycles are required
even at good and typical weather conditions.
The capability for simultaneous dual frequency observations
(T$^\nu_{\rm swt}$=0) achieves a perfect tropospheric calibration at 
any frequency, eliminating the need for fast switching, 
extending the application of SFPR toward the very high frequency domain.
Our analytical studies show its feasibility for observations at 43/86 GHz, 
and the trend shown in Figure \ref{fig1}c) 
will continue to be the same at higher frequencies. 
Therefore, using simultaneous dual frequency SFPR observations 
allows one to achieve the full potential accuracy of the very
precise VLBI phase observable for astrometric measurements, 
and enables long coherence times under all weather
conditions, for both ground and S-VLBI observations even at high
frequencies. 
Hence, we strongly recommend the inclusion of simultaneous dual
frequency capability in the mission specifications for future S-VLBI, 
especially at mm-wavelengths.

\subsection{Extrapolation of our results to other regions of parameter space}

Our SFPR S-VLBI simulations mainly use the specifications of the ASTRO-G
satellite antenna as a starting point, and test a region of the
parameter space as described in Table \ref{tab:par}. 
Here we discuss the extrapolation 
of these results to other missions and regions in the parameter space,
namely higher frequencies, different orbits and orbit errors,
and aim at extracting information of general interest for S-VLBI. 
Missions such as RadioAstron, Millimetron and others have been proposed 
to operate in these domains. 

In SFPR observations with RadioAstron at C/K-bands
the dominant ionospheric errors are expected to increase with the pair angular
separation.  In this case, occasional observations of a nearby
calibrator, $\le 1^\circ$ away, along with the target source are recommended.
The astrometric accuracy is expected to increase if ionospheric errors 
are kept small.
Other than astrometry, we forsee a useful application of FPT observations of the
target source to alleviate sensitivity issues at K-band, arising from the expected 
decrease in the correlated fluxes observed with such a large orbit, using 
simultaneous observations at C-band to increase the coherence time.
Our simulations show the capability of this technique to compensate
for errors in the orbit determination.
The only requirement on the a-priori orbit determination is set by the
delay and rate windows in the correlator processing,
for which measurements with errors of a few meters, 
achievable with conventional satellite tracking techniques, are suitable.
Hence, SFPR (and FPT) with RadioAstron, even though $\nu^{low}$ is 
below the suggested range, will produce significant benefits.

S-VLBI observations up to very high frequencies ($\nu^{high}\ge$ 43 GHz) should be
feasible using simultaneous dual frequency observations,
even for objects that are too weak to be directly detected, 
provided they can be detected at a lower frequency $\nu^{low}$. 
In order to minimize the impact of the ionospheric residual terms, it is
recommended to use as a reference frequency $\nu^{low} \sim$15--22
GHz. Intermittent  observations, up to tens of minutes, of a distant calibrator, up to 
several degrees away, should be suitable.   
The increase in resolving power will result
in a parallel increase in the astrometric accuracy, particularly if
perfect tropospheric compensation is obtained using simultaneous dual
frequency observations, and assuming reasonable SNR values.  
A study which addresses the case of weak sources and complex structure
will be performed in a future series of simulations.
In addition, we do not foresee a major impact on the relative
performance trends obtained from our simulations.
The limit on the highest frequency is not set by the
SFPR requirements, but most likely by the surface accuracy required
for the satellite antenna. 
Hence, having simultaneous dual frequency observations improves
the performance in all cases, and at frequencies higher than 43 GHz
this capability is an imperative.

Ground PR observations beyond 43 GHz are also limited by the rapid 
tropospheric fluctuations. Here too the use of SFPR techniques 
would extend the benefits currently achieved with PR to a much 
higher frequency domain. 
An early demonstration of the dual frequency calibration step for  
connected interferometry observations at 19/146 GHz can be found in  
\citet{asaki_98}.
Application of the two step SFPR technique to sub-mm VLBI observations 
with Atacama Large Millimeter/submillimeter Array (ALMA), for example, is an area of great interest. We are investigating the
considerations and requirements for SFPR at the highest frequencies,
also in comparison with Water Vapor Radiometer (WVR) phase corrections, 
and these will be discussed in future publications.

\subsection{Outcomes from combining SFPR and PR techniques} 
The SFPR and PR techniques are complementary in providing astrometric
measurements and detection of weak sources in a wide frequency
domain and are widely applicable. The direct outcome of SFPR techniques are
measurements of the 
relative separation between the emitting regions at the two observed frequencies, 
even at mm-wavelengths; this is of direct interest for 
studies of the core-shift phenomena in AGNs, 
the alignment of spectral line emission -- for
example SiO masers at different transitions or from different
molecules such as H$_2$O and SiO -- but in general to any position shift
regardless of its origin.  A07 have shown there is reasonable
probability of success of PR observations with S-VLBI for frequencies 
up to 22 GHz.  The combination of astrometric
results from SFPR between $\nu^{high}$ and $\nu^{low}$, with
conventional PR at the lower frequency $\nu^{low}$, results in
astrometric measurements with respect to an external source at
$\nu^{high}$, even though conventional PR at $\nu^{high}$ may not be
feasible.  Therefore, the fields of application
mentioned above
extend to studies that require the comparison of positions at different epochs,
such as proper motion and parallax studies, or stability studies,
at the highest frequencies.  
Hence, the combination of both techniques opens a much larger scope of
application, even at high frequencies, with ground and S-VLBI.

\section{Conclusions}

To summarize, this paper investigates the SFPR technique applied to S-VLBI, to achieve astrometry and
increased coherence time for detection of weak sources. Our
comprehensive simulation and analytical studies comprise observations
between 5 and 86 GHz, 
either alternated with a switching cycles of 1 and 2 minutes, 
or simultaneously observed, of pairs of sources with
separations up to several degrees, source switching cycles up to many
minutes, orbit errors larger than 1 meter, and coherence studies,
under all weather conditions.  
A list of the more significant results obtained from this study is:
\begin{enumerate}
\item SFPR solves the specific issues of S-VLBI that
  limit the application of conventional phase referencing techniques;
\item Our simulations show that $\mu$-arcsecond level astrometry at 43 GHz
  can be achieved in most cases, using source pair angular separations up
  to several degrees, and slow source switching cycles with S-VLBI;
\item The satellite antenna orbit error is readily compensated,
  along with any other non dispersive error, with observations at two
  frequencies. Hence conventional satellite tracking techniques are sufficient for
  orbit reconstructions;
\item The same applies to the tropospheric fluctuations which are readily
  compensated, and result in long coherence times up to several hours.
  Hence weaker sources can be targeted;
\item The capability for simultaneous dual frequency SFPR observations with S-VLBI is 
  mandatory at $\nu^{high}\ge 43$ GHz, improves the performance at any frequency, and 
  eliminates fast frequency switching operations. Hence this is 
  a crucial capability to include in future mission specifications. At 22/43 GHz,
  frequency switching cycles faster than approximately 0.4 and 1 minutes, for poor and other
  weather conditions, respectively, are required;
\item RadioAstron specifications are compatible with SFPR observations, 
 with $\nu^{low}$ of 5 GHz. They are expected to be dominated by ionospheric errors, 
 hence a small source pair angular separation ($\sim 1^\circ$) is recommended.
\item Ground VLBI also would benefit from simultaneous dual frequency
  capability at the highest frequencies observed, 
  especially for compensating tropospheric fluctuations and 
  enabling detection of weaker sources.
\end{enumerate}

\section{Acknowledgements}

The authors acknowledge support by The University of Western Australia (UWA) to complete this publication,
through the Research Collaboration Awards 2009 Round 2 funding program for the
project ``Enabling state-of-the-art Astrometry with the Japanese Space Mission VSOP-2''.


\end{document}